\documentclass[ prl, twocolumn, amsfonts, amsmath,floatfix] {revtex4-1}
\pdfoutput=1
\usepackage{graphicx}
\usepackage{amssymb,amsmath}
\usepackage{xspace}
\usepackage[usenames, dvipsnames]{color}
\usepackage{soul}
\usepackage{lipsum}
\usepackage{natbib}
\usepackage[mathscr]{eucal}
\usepackage{dsfont}

\renewcommand{\vec}[1]{{\mathbf #1}}
\newcommand{\av}[1]{\langle #1\rangle}
\newcommand{\ket}[1]{\ensuremath{|#1\rangle}\xspace}
\newcommand{\bra}[1]{\ensuremath{\langle #1|}\xspace}

\newcommand{\be}{\begin{equation}}
\newcommand{\ee}{\end{equation}}

\usepackage[usenames,dvipsnames]{color}

\begin{document}

\title{Optimizing light storage in scattering media with the dwell-time operator}

\author{M. Durand, S. M. Popoff, R. Carminati, and A. Goetschy
}

\affiliation{
ESPCI Paris, PSL University, CNRS, Institut Langevin, 1 rue Jussieu, F-75005 Paris, France
}

\begin{abstract}
We prove that optimal control of light energy storage in disordered media can be reached by wavefront shaping. For this purpose, we build an operator for dwell-times from the scattering matrix, and characterize its full eigenvalue distribution both numerically and analytically in the diffusive regime, where the thickness $L$ of the medium is much larger than the mean free path $\ell$. We show that the distribution has a finite support with a maximal dwell-time larger than the most likely value by a factor $(L/\ell)^2\gg 1 $. This reveals that the highest dwell-time eigenstates deposit more energy than the open channels of the medium. Finally, we show that the dwell-time operator can be used to store energy in resonant targets buried in complex media. 
\end{abstract}

\maketitle

Recent developments in wavefront shaping protocols have allowed spectacular demonstrations of light manipulation in complex media~\cite{mosk12, rotter17}, such as non-invasive imaging in biological tissues~\cite{horstmeyer15, yu15}, focusing~\cite{vellekoop08} or enhanced power delivery~\cite{ kim12, popoff14, hsu17} behind opaque media, or focusing~\cite{choi13, ambichl17, katz19} and enhanced absorption~\cite{liew16} inside scattering materials. The large number of degrees of freedom supported by disordered systems has also been proposed as a ressource for imaging with high resolution~\cite{vellekoop10, park14}, controlling the strength of light-matter interaction~\cite{bachelard14, davy18, pichler19}, or performing optically complex and reconfigurable operations~\cite{matthes19}. 

In this context, great attention has been given to the statistical properties of the transmission matrix $t$~\cite{popoff10a, shi12, goetschy13_prl, gerardin14, hsu15}. This matrix admits a significant fraction of singular states, called open channels, responsible for complete destructive (constructive) interference in reflection (transmission), even if the medium is opaque on average~\cite{dorokhov84, nazarov94}. The intensity map inside the medium resulting from the propagation of open channels has also been elucidated~\cite{davy15a}, revealing a bell-shape profile along the propagation direction very different from the familiar linear decay obtained with plane wave illumination~\cite{vanrossum99}. This property makes open channels good candidates for enhancing significantly energy deep inside disordered media~\cite{sarma16, koirala17, hong18}. However, these states are, by construction, those maximizing the output flux (they are eigenstates of the operator $t^\dagger  t$ associated with the largest eigenvalues), and not necessarily the stored energy. The transverse localization of open channels, discovered very recently~\cite{yilmaz19}, also supports the idea that they are not necessarily the ones optimizing  energy storage. 

In this Letter, we explicitly build an operator for the dwell time (or stored energy) in complex media illuminated with monochromatic light. Its expression can be obtained directly from the scattering matrix (including evanescent channels) of the disordered material and the dispersion properties of the surrounding medium. First, we show that the dwell-time (DT) operator is not strictly identical to the Wigner-Smith matrix, introduced historically to characterize the duration of a scattering process~\cite{smith60}, by identifying a contribution resulting from the interference between the incident and scattered fields, similar to that predicted in 1D for electrons~\cite{winful03}. Second, we study its  eigenvalue distribution $p(\tau)$ for wave propagating through a disordered slab of thickness $L\gg \ell$, where $\ell$ is the light mean free path. We find that for non-resonant scattering this distribution is parametrized by two time scales only: the scattering time $\tau_s\sim \ell/c$ ($c$ being the speed of light in vacuum),  as well as the mean time $\av{\tau}\sim L/c$, which is known to be remarkably independent of the disorder strength~\cite{pierrat14, savo17}. It also exhibits a dominant peak at $\tau\sim \tau_s$, and has a finite support with a maximal DT eigenvalue $\tau^{\textrm{max}} \sim \av{\tau}^2/\tau_s$. This last result implies that the maximal energy that can be stored in a disordered medium by wavefront shaping with fixed input power $\phi^{\textrm{in}}$ scales as $U^{\textrm{max}}\sim \phi^{\textrm{in}} \tau_{\textrm{Th}}$, where $\tau_{\textrm{Th}}$ is the Thouless time. Finally, we also show that the DT operator is a powerful tool to selectively deposit energy on local resonant targets embedded in a given realization of a complex medium.

Let us start with the construction of the DT operator. For clarity, we restrict  the present discussion to the propagation of scalar waves in non-resonant and non-absorbing materials, described by the equation $\left[\nabla^2 + k^2\epsilon(\vec{r}) \right]\psi(\vec{r})=0$. Here, $\epsilon(\vec{r})$ is the (real) dielectric function, and $\psi$ is the complex amplitude of the monochromatic wave with frequency $\omega=ck$. The quantity to maximize is the electromagnetic energy $U=\epsilon_0 \int_{\mathcal{V}}\textrm{d}\vec{r}\, \epsilon(\vec{r})\vert\psi(\vec{r})\vert^2/2$, where $\mathcal{V}$ is the volume occupied by the  disordered slab. From the wave equation, we readily obtain the relation
\be
\epsilon(\vec{r})\vert\psi(\vec{r})\vert^2
=\frac{c^2}{2\omega}
\mathbf{\nabla} .
\left(\partial_\omega \psi \nabla \psi^*
-\psi^*\partial_\omega  \nabla \psi
\right),
\ee
which allows us to express the energy as the surface integral
\be
U=\frac{\epsilon_0c^2}{4\omega}\int_{\mathcal{S}}
\textrm{d}\vec{r}\,\vec{n}. 
\!\left(\partial_\omega \psi \nabla \psi^*
-\psi^*\partial_\omega  \nabla \psi
\right),
 \label{U}
\ee
 where $\mathcal{S}$ denotes the input and output surfaces of the slab and $\vec{n}$ the outward normal on them.
For a  Schr\"{o}dinger wave $\psi$, a similar relation holds for the probability  $\int_{\mathcal{V}}\textrm{d}\vec{r}\, \vert\psi(\vec{r})\vert^2$, which is independent of the explored potential~\cite{smith60},  contrary to $U$ that depends on  $\epsilon(\vec{r})$.
Next, we express the field $\psi$ on each surface in terms of the incident field $\psi^{\textrm{in}}$, and the reflection and transmission matrices. Care must be taken since the scattered field at the sample surface can have contributions from both propagating and evanescent channels of the surrounding medium. 
Here, we consider a disordered slab embedded in a multimode waveguide supporting $N$ propagating channels, so that the reflection and transmission matrices restricted to this channel basis, noted $r$ and $t$ respectively,  are of size $N\times N$. Relegating the technical derivation to the Supplementary Material (SM)~\cite{SI}, we find that the stored energy can be expressed as the expectation value
\be
U=\phi^{\textrm{in}}\bra{\psi^{\textrm{in}}}Q_d\ket{\psi^{\textrm{in}}},
 \label{U2}
\ee
where the DT operator $Q_d$ is a sum of three contributions, with clear and distinct meanings discussed below:
 \be
 Q_d=Q+Q_e+Q_i.
 \label{Qd}
 \ee
Depending on the physical situation of interest (choice of $\psi^{\textrm{in}}$, size and scattering strength of the medium), each of these terms can produce the dominant contribution to $U$.  We discuss their expressions below for an incident field without evanescent component, and injected from one side of the slab only, which corresponds to the most common experimental situation. More general expressions are given in the SM~\cite{SI}.
 
The first term in the right-hand side in Eq.~(\ref{Qd}) is the well-known Wigner-Smith matrix $Q=-\textrm{i}(t^\dagger\partial_\omega t+r^\dagger\partial_\omega r)$, which characterizes the duration of the scattering process for quasi-monochromatic signals measured in the far field of the medium~\cite{smith60, texier16, rotter17}. The utility of this operator for controlling wave propagation in multimode fibers and scattering media has been demonstrated in recent years~\cite{rotter11, carpenter15, davy15b, gerardin16, xiong16, ambichl17}.   
The second term captures scattering contributions into evanescent channels, and contributes even if $\psi^{\textrm{in}}$ has no evanescent component.
It reads $Q_e=(t_e^\dagger D^e t_e+r_e^\dagger D^e r_e)/2$, where $r_e$ and $t_e$ are the reflection and transmission matrices into evanescent channels of the waveguide~\cite{mello04}. The matrix $D^e$ is diagonal, with elements $D^e_{\alpha \beta}=\frac{\partial_\omega \kappa_\alpha}{\kappa_\alpha}\delta_{\alpha \beta}$, where $\kappa_\alpha=\sqrt{q_\alpha^2-k^2}$ is the inverse decay length of the evanescent channel $\alpha$  ($q_\alpha=\alpha \pi /W$ in a 2D waveguide of width $W$). The contribution of $Q_e$ cannot be neglected close to the onset of a new propagating mode of the waveguide~\cite{bagwell90, gomez01, SI}. The important impact of evanescent channels on dwell-times has also been revealed recently in the case of scattering by subwavelength particles~\cite{shen19}. However, the contribution of $Q_e$ to the distribution $p(\tau)$ studied below turns out to be negligible for all frequencies except a discret set (see SM), and will not be discussed further. 

The third term in the decomposition~(\ref{Qd}) describes a qualitatively different contribution, due to the interference between the incident and reflected propagating fields. Since the total field is $\psi \sim \psi^{\textrm{in}}+r\psi^{\textrm{in}}$ on the front surface and  $\psi \sim t\psi^{\textrm{in}}$ on the back surface, the field products in Eq.~(\ref{U}) involve cross-terms in reflection only. The associated matrix reads $Q_i=-\textrm{i}(Dr-r^\dagger D)/2$, where $D$ has elements $D_{\alpha \beta}=\frac{\partial_\omega k_\alpha}{k_\alpha}\delta_{\alpha \beta}$, and $k_\alpha=\sqrt{k^2-q_\alpha^2}$. In the SM~\cite{SI}, we provide an alternative proof of Eq.~(\ref{Qd}) based on the continuity equation, that highlights the close connection between $Q_i$ and the interference term between the incident and scattered field at the origin of the optical theorem~\cite{akkermans07}. The contribution of $Q_i$ becomes appreciable for states with large reflection. For this reason, it cannot be neglected in strongly scattering media that reflect most of the light. In particular, it contributes substantially to the lower part of the DT eigenvalue distribution in the regime $L\gg \ell$ (see SM for details). 

Equation~(\ref{Qd}) generalizes to arbitrary scattering media and arbitrary dimension (arbitrary $N$) the relation established for electrons in the case $N=1$~\cite{winful03}, or for electromagnetic waves~\cite{winful03b}, interacting with a simple barrier (for which $Q_e$ is zero). For electrons, the trace of the matrix $Q_i$ is known as a correction to the Friedel sum rule~\cite{texier16}, that relates the density of states ($\sim \textrm{Tr}Q_d$) to the Wigner time delay ($\sim \textrm{Tr}Q$). In its operator form, the difference between $Q_d$ and $Q$ has also been pointed in Refs.~\cite{lagendijk96, sokolov97, rotter17}, but not expressed in the explicit and computationally useful expansion given by Eq.~(\ref{Qd})~\cite{ambichl}. 

\begin{figure}[t]
\includegraphics[width=0.9\linewidth]{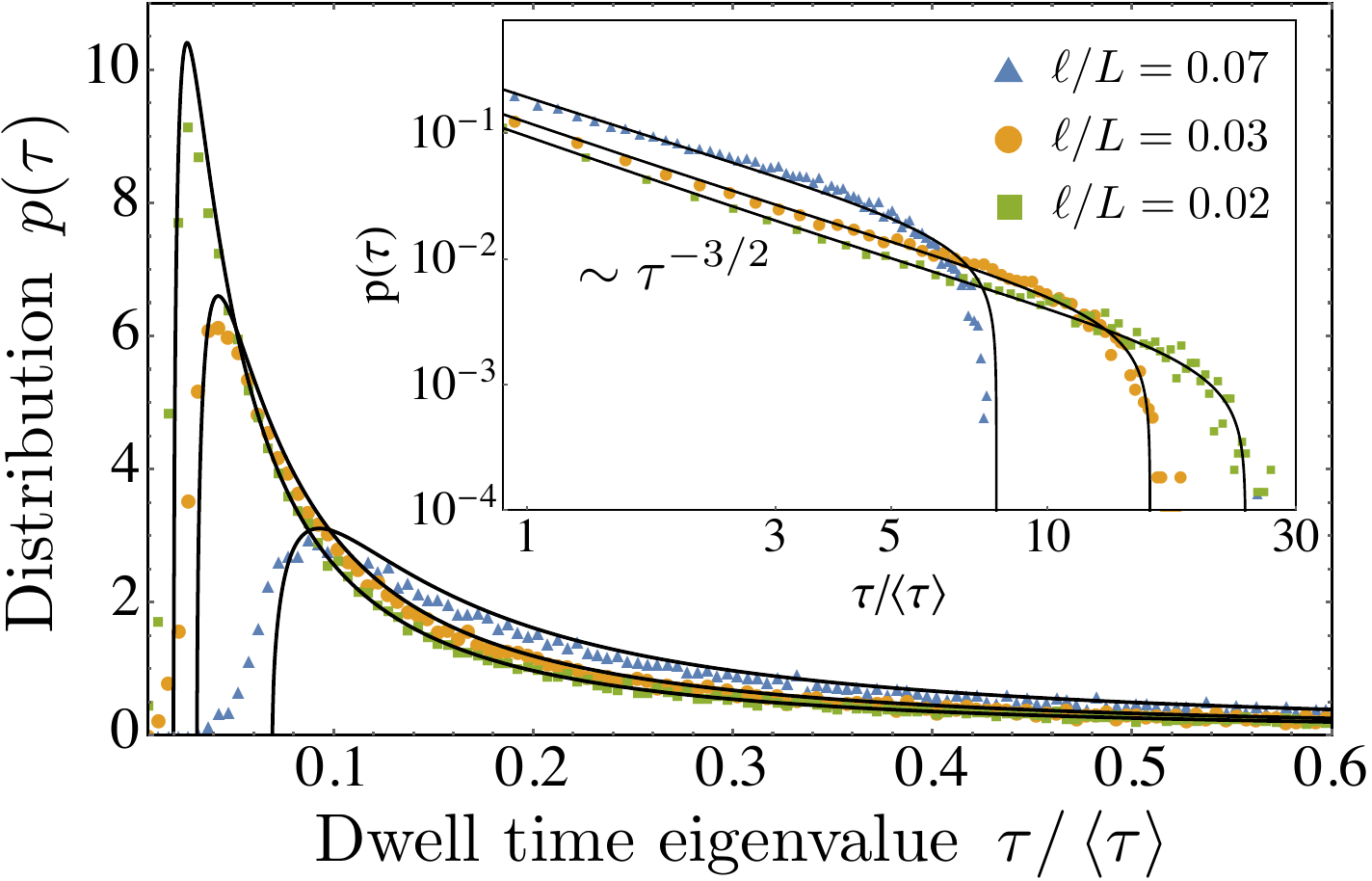}
\caption{ Eigenvalue distribution of the dwell-time operator $Q_d$, evaluated for a disordered slab (length $kL=300$) embedded in a multimode waveguide ($N=287$). Analytical predictions (solid lines) are compared with numerical results (dots) obtained from the solution of the wave equation for $128$ realizations of the slab, with dielectric function $\epsilon(\vec{r})=n_1^2+ \delta \epsilon(\vec{r})$; $n_1=1.5$ and $\delta \epsilon(\vec{r})$ is uniformly distributed in $[-a, a]$ for each discretization cell. Results for three values of $a$ are represented, corresponding to $k\ell=21.4, \,9.3,\, 5.8$.
}
\label{fig1}
\end{figure} 

\begin{figure*}[t]
\includegraphics[width=0.8\linewidth]{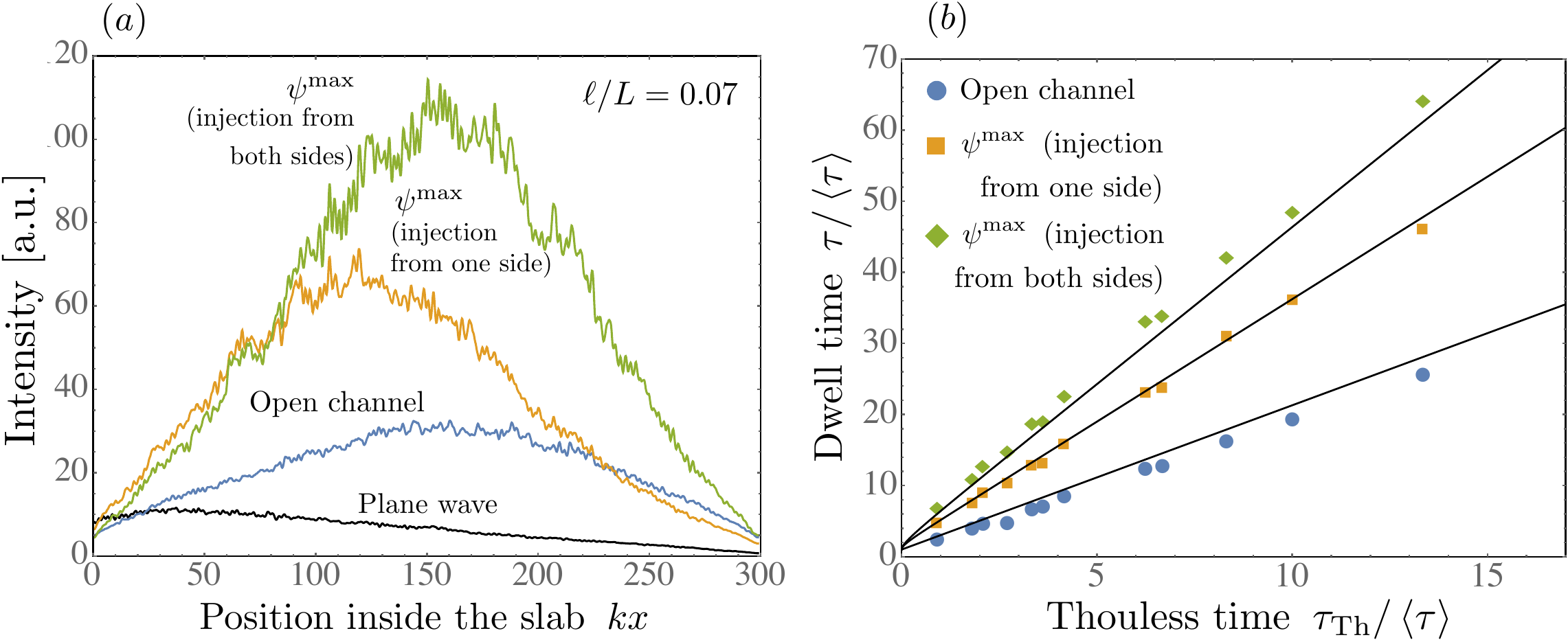}
\caption{(a) Intensity profiles inside a disordered slab (integrated over the transverse dimension) resulting from the propagation of different input states $\psi^{\textrm{in}}$: the first mode of the waveguide (similar to a plane wave), the most open channel, and the eigenstate $\psi^{\textrm{max}}$ of $Q_d$ associated to the largest eigenvalue $\tau^{\textrm{max}}$. 
(b) Dwell times $\tau = \bra{\psi^{\textrm{in}}}Q_d \ket{\psi^{\textrm{in}}}$, averaged over $128$ configurations (dots), corresponding to the different states $\psi^{\textrm{in}}$ shown in (a), versus the Thouless time $\tau_{\textrm{Th}}=L^2/\pi^2D_B=  2L^2/\pi^2\ell v$. All times are normalized by  $\av{\tau}= (\pi/2)L/v$. Solid lines correspond to analytical predictions (see text for details). 
 } 
\label{fig2}
\end{figure*} 

To characterize the properties of the matrix $Q_d$, we performed extensive numerical simulations of scalar wave propagation through two-dimensional disordered slabs placed in a multimode waveguide, using the recursive Green's function method~\cite{goetschy13_prl}, and computed $Q_d$ as defined in Eq.~(\ref{Qd}). The eigenvalue distribution $p(\tau)$ of $Q_d$ is represented in Fig.~\ref{fig1} for three values of the disorder strength $1/k\ell$, in the diffusive regime $kL\gg k\ell \gg 1 $. We find that $p(\tau)$ has a pronounced peak that shifts towards small time as $k\ell$ decreases. This illustrates the fact that most of the light experiences a few scattering events before being reflected after a time $\sim \ell/c$. On the other hand, a close look at the largest eigenvalues (see inset) reveals that  $p(\tau)$ is bounded, with an upper edge $\tau^{\textrm{max}}$ that increases with the disorder strength. This effect is triggered by light crossing the sample by diffusion, a process that takes more time when the mean free path is reduced since the number of scattering events is increased. 

To support the previous observations, we developed an analytical model for the distribution $p(\tau)$. Since $Q_d$ and $Q$ have similar spectra, with differences observed at small times only (see SM), we work with $Q$ in the theoretical development.   
First, following Ref.~\cite{beenakker01}, we use the simple relation that links the scattering matrix $S=\left(\begin{smallmatrix}r\\t\end{smallmatrix}\right)$ in absence of absorption, to the scattering matrix $S^a(\omega)=S(\omega+\textrm{i}/2\tau_a)$ with uniform absorption, where $\tau_a$ is the absorption time. A first-order expansion gives $Q(\omega)\simeq\tau_a [1-S^a(\omega)^\dagger S^a(\omega)]$ for $\omega \tau_a\gg1$. The advantage of this relation lies in the fact that the joint probability distribution (JPD) of the eigenvalues of the operator $S^a(\omega)^\dagger S^a(\omega)$ is known, for disordered media excited from one side, in the limit $L\to \infty$~\cite{bruce96, beenakker96}. Denoting by $\tau_n$ the eigenvalues of $Q$, the JPD of the decay rates $\gamma_n=1/\tau_n$ takes the form of the Gibbs distribution $p(\{\gamma_n \})\sim e^{-\mathcal{H}}$, with
\be
\mathcal{H}=2N\tau_s\sum_{n=1}^N\gamma_n-\sum_{n<m}\textrm{ln}\vert \gamma_n - \gamma_m\vert,
\label{jpdfGamma}
\ee
where $\tau_s$ is the scattering time. In 2D, for non-resonant scattering, it has to be defined as $\tau_s=(\pi/2)\ell/v$, where $v=c/n$ is the group velocity, $n$ being  the effective refractive index of the disordered slab (see SM). The Laguerre distribution defined by Eq.~\eqref{jpdfGamma} implies that the matrix $Q^{-1}$ behaves as a Wishart matrix in a disordered medium, a property which is also true in a chaotic cavity~\cite{brouwer97, grabsch16}. In the first case, $Q^{-1}$ has the same JPD as the matrix $HH^\dagger/\tau_s$, where $H$ is a $N\times N$ complex Gaussian random matrix, while in the second case, $Q^{-1}$ has the JPD of $HH^\dagger/\left<\tau \right>$, where $H$ is of size $N\times N/2$.

The result in Eq.~(\ref{jpdfGamma}) was obtained for infinite-size systems. In this limit, the marginal distribution $p(\tau)$ depends on $\tau_s$ only, with infinite $\av{\tau}=\left<\sum_{n=1}^N\tau_n \right>/N$. In non-resonant systems of finite size $L$, it is known that $\av{\tau}$ scales as $\sim L/c$, both in the quasi-ballistic and diffusive regimes~\cite{pierrat14, savo17}; in 2D, $\av{\tau}= (\pi/2)L/v$. To restore a finite mean time, we make the ansatz that $\mathcal{H}$ is still well approximated by expression~(\ref{jpdfGamma}) in the regime $L/\ell\gg1$ and $N\gg 1$, and search for the marginal distribution $p(\tau)$ that minimizes $\mathcal{H}$ under the constraint $\int \textrm{d}\tau p(\tau)=\av{\tau}$. We find (see the SM for details)
\be
p(\tau)\simeq \frac{2\tau_s}{\pi\tau^2}\sqrt{ \left( \alpha\frac{\tau}{\tau_s}-1\right) \left(1-\beta\frac{\tau}{\tau_s}\right)}\left(1+\gamma \frac{\tau}{\tau_s}\right),
\label{poftau}
\ee
where $\alpha$, $\beta$, and $\gamma$ obey a set of three coupled equations depending on the single parameter $b\equiv\av{\tau}/\tau_s=L/\ell$. At large optical thickness ($b\gg1$), they reduce to $\alpha \simeq 1$, $\beta \simeq 9/[4b(b-4)]$, and $\gamma \simeq 2\beta$. The distribution~(\ref{poftau}) is maximum for $\tau\simeq4\tau_s/3$, and has a finite support $[\tau^{\textrm{min}}, \tau^{\textrm{max}} ]$, with $\tau^{\textrm{min}}\simeq\tau_s$ and
\be
\tau^{\textrm{max}}\simeq\frac{4}{9}\av{\tau}\left(\frac{\av{\tau}}{\tau_s} -4 \right).
\label{tauMax1}
\ee
These theoretical results are in excellent agreement with the numerical simulations, without adjustable parameter, as shown in Fig.~\ref{fig1}. The distribution also exhibits a power $p(\tau)\sim\tau^{-3/2}$, for $\av{\tau} \lesssim \tau \lesssim \tau^{\textrm{max}}$. This is a hallmark of diffusion, also observed in numerical simulation of the 2D kicked rotor dynamics~\cite{ossipov03}.
 
Our analysis reveals the existence of a finite maximal eigenvalue $\tau^{\textrm{max}}$ of $Q^d$. According to Eq.~\eqref{U2}, the corresponding eigenstate,  $\ket{\psi^{\textrm{max}}}$, should yield the largest amount $U^{\textrm{max}}$ of stored energy. To check this prediction, we compared the intensity pattern produced  inside the slab by $\ket{\psi^{\textrm{max}}}$ with that resulting from the propagation of other remarkable wavefronts, such as the most open channel $\ket{\psi^{\textrm{oc}}}$ (the eigenstate of $t^\dagger t$ associated to the largest transmission eigenvalue $T\simeq 1$). Representative results are shown in Fig.~\ref{fig2}(a). Contrary to the intensity profile created by the first mode of the waveguide (which behaves as a plane wave), both $\ket{\psi^{\textrm{oc}}}$ and $\ket{\psi^{\textrm{max}}}$ give rise to a concentration of energy deep inside the medium. In addition, the intensity pattern due to  $\ket{\psi^{\textrm{max}}}$ is significantly larger than that due to $\ket{\psi^{\textrm{oc}}}$, when integrated over the transverse dimension $y$. This clearly shows that $U^{\textrm{max}}>U^{\textrm{oc}}$.

Using the analytical result~\eqref{poftau}, the ratio $U^{\textrm{max}}/U^{\textrm{oc}}$ can be evaluated precisely. In Ref.~\cite{davy15a}, it was shown that the intensity profile created by $\ket{\psi^{\textrm{oc}}}$ in 2D is $I^{\textrm{oc}}(x)=\int\textrm{d}y\vert\psi^{\textrm{oc}}(x,y)\vert^2 =(\pi/2)[1+(\pi/2) (L/\ell)x(L-x)/L^2]/k$, where $x$ is the direction perpendicular to the slab. After integration over $x$, we obtain $U^{\textrm{oc}}=\phi^{\textrm{in}}\tau^{\textrm{oc}}$, with
\be
\tau^{\textrm{oc}}\simeq\frac{\pi}{12}\av{\tau}\left(\frac{\av{\tau}}{\tau_s} +\frac{12}{\pi} \right).
\label{tauOC}
\ee
Hence, $U^{\textrm{oc}}$ grows quadratically with the sample thickness $L$. It is much larger than the energy $U^{\textrm{pw}}$ deposited by a plane wave, that grows linearly with $L$ [$I^{\textrm{pw}}(x)\sim (L-x)/(kL)$ gives $U^{\textrm{pw}}\sim \av{\tau} \phi^{\textrm{in}}$], but it is smaller than $U^{\textrm{max}}$. 
Indeed, Eqs.~(\ref{tauMax1}) and~(\ref{tauOC}) give $U^{\textrm{max}}/U^{\textrm{oc}}\simeq 16/3\pi>1$.
These predictions are confirmed by the results of numerical simulations presented in Fig.~\ref{fig2}(b), where $\tau^{\textrm{max}}=U^{\textrm{max}}/\phi^{\textrm{in}}$ and $\tau^{\textrm{oc}}=U^{\textrm{oc}}/\phi^{\textrm{in}}$ are shown to be both larger than the Thouless time, $\tau_{\textrm{Th}}=L^2/(\pi^2D_B)$, which is the longest mode lifetime of the diffusion equation. In 2D, the light diffusion constant is $D_B=\ell v/2$, so that $\tau^{\textrm{max}}\simeq(\pi^3/9)\tau^{\textrm{Th}}$ and $\tau^{\textrm{oc}}\simeq(\pi^4/48)\tau^{\textrm{Th}}$ for $L\gg\ell$. 
In Fig.~\ref{fig2}(b), we also show that the ratio  $\tau^{\textrm{max}}/\tau^{\textrm{Th}}$ can be  increased by injecting light from both sides of the sample. In this case, the average intensity profile $I^{\textrm{max}}(x)$ presents a mirror symmetry with respect to the middle of the slab $x=L/2$, as imposed by statistical invariance [see Fig.~\ref{fig2}(a)]. Inspired by the microscopic approach developed in Ref.~\cite{ossipov18}, we could establish (see SM for details) an expression for  $\tau^{\textrm{max}}$ in this situation, which reads
\be
\tau^{\textrm{max}}\simeq \tau_s
 \left[
 \zeta
 \left( \frac{\av{\tau}^2}{\tau_s^2} +4\frac{\av{\tau}}{\tau_s} -4\right) -1
  \right]
\label{tauMax2}
\ee
at large optical thickness, where $\zeta\simeq 0.57$ is the solution of a transcendental equation. This prediction also agrees with numerical simulations [see Fig.~\ref{fig2}(b)].

\begin{figure}[t]
\includegraphics[width=0.95\linewidth]{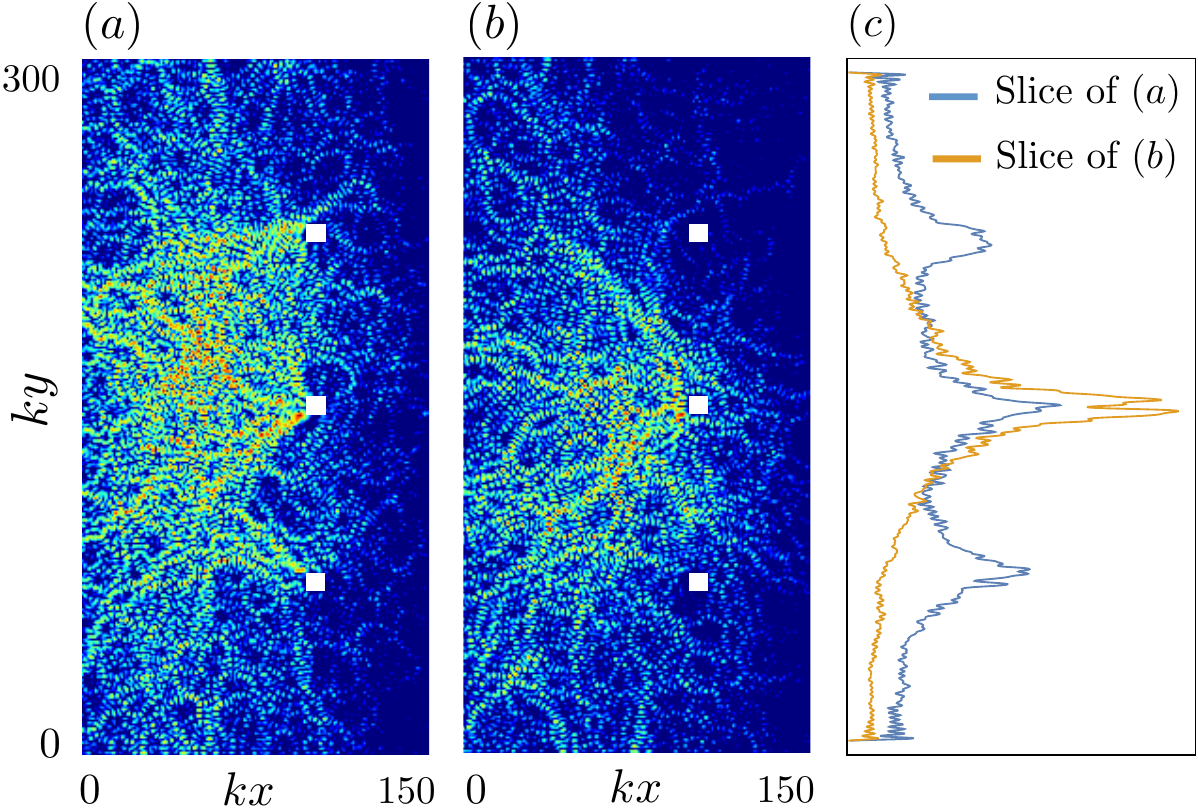}
\caption{ 
(a) and (b) Typical intensity patterns $I(x,y)$ resulting from the propagation of  the states associated with the maximal eigenvalue of the absorption operator (a) and the extremal eigenvalue of $Q_d^H$ (b), in a disordered medium ($kL=150$, $k\ell=21.4$) containing three absorbers placed at depth $kx=100$ (white squares). Only the second absorber is resonant ($\mathcal{F}=10^{-4}$, $\omega_0/\Gamma=10^4$). 
(c) Slices $I(y) = \int_{80/k}^{95/k}\textrm{d}x I(x,y)$ of the patterns (a) and (b), averaged over $32$ configurations, reveal the different performances of $A$ and $Q_d^H$ in terms of storage inside the absorbers. 
}
\label{fig3}
\end{figure} 

We have discussed the properties of the DT operator $Q_d$ in statistically homogeneous non-resonant disordered materials, and demonstrated quantitatively superior (yet qualitatively similar) performances of the largest DT eigenstates for energy storage, compared to open channels. However, $Q_d$ is an operator specifically constructed to optimize the quality factor of an arbitrary complex structure. This concept is radically different from the monochromatic scattering properties usually captured by matrices such as $t^\dagger t$ or $r^\dagger r$. To illustrate this last point, let us consider a set of small absorbers buried in an otherwise non-absorbing disordered medium. One of them, resonant at frequency $\omega_0$ with quality factor $\omega_0/\Gamma\gg 1$, is described by the Lorentzian dielectric function $\epsilon = 1 +\omega_0^2\mathcal{F}/(\omega_0^2-\omega^2-i\omega \Gamma)$. Our goal is to compare the performances of the absorption matrix $A=1-t^\dagger t-r^\dagger r$ and the DT matrix. In the presence of absorption, we can show that the DT matrix is related to the Hermitian part of $Q_d$ defined as $Q^H_d=(Q_d+Q_d^\dagger)/2$ (see SM). We show in Fig.~3 the intensity patterns calculated inside the medium at the resonance frequency $\omega_0$, and resulting from the propagation of the states $\ket{\psi^{a}}$, associated to the largest absorption eigenvalue, and $\ket{\psi^{H}}$, associated to  the extremal eigenvalue of $Q^H_d$. We clearly see that $\ket{\psi^{a}}$ deposits energy indistinctly on all absorbers, whereas $\ket{\psi^{H}}$ focuses specifically on the resonant scatterer that will induce the largest dwell time.

In summary, we have presented a general setting for tuning light storage in arbitrary complex media based on the dwell-time operator~\eqref{Qd}. We showed that the eigenvalue  distribution of this operator takes the universal form~\eqref{poftau} in the diffusive regime,
and demonstrated the possibility to reach more than one hundred percent energy increase with respect to what can be achieved with open channels. Finally, we established  that $Q_d$ can be used for  addressing hidden resonant targets without need for guide stars.

\begin{acknowledgments}
We thank R. Pierrat for his help in the implementation of the recursive Green's function method at the early stage of the work. This work was supported by LABEX WIFI (Laboratory of Excellence within the French Program Investments for the Future) under references ANR-10- LABX-24 and ANR-10-IDEX-0001-02 PSL*.
\end{acknowledgments}

%

\end{document}


\title{Supplementary Material \\
Optimizing light storage in scattering media with the dwell-time operator}

\author{M. Durand, S. Popoff, R. Carminati, and A. Goetschy
}

\affiliation{
ESPCI Paris, PSL  University, CNRS, Institut Langevin, 1 rue Jussieu, F-75005 Paris, France
}

\maketitle

\section{I. Proof, generalization and interpretation of formula (3) and (4) of the main text}

\subsection{I.1 Case 1: $\epsilon(\vec{r})$ real and independent of frequency} 
\begin{figure}[t]
\includegraphics[width=0.8\linewidth]{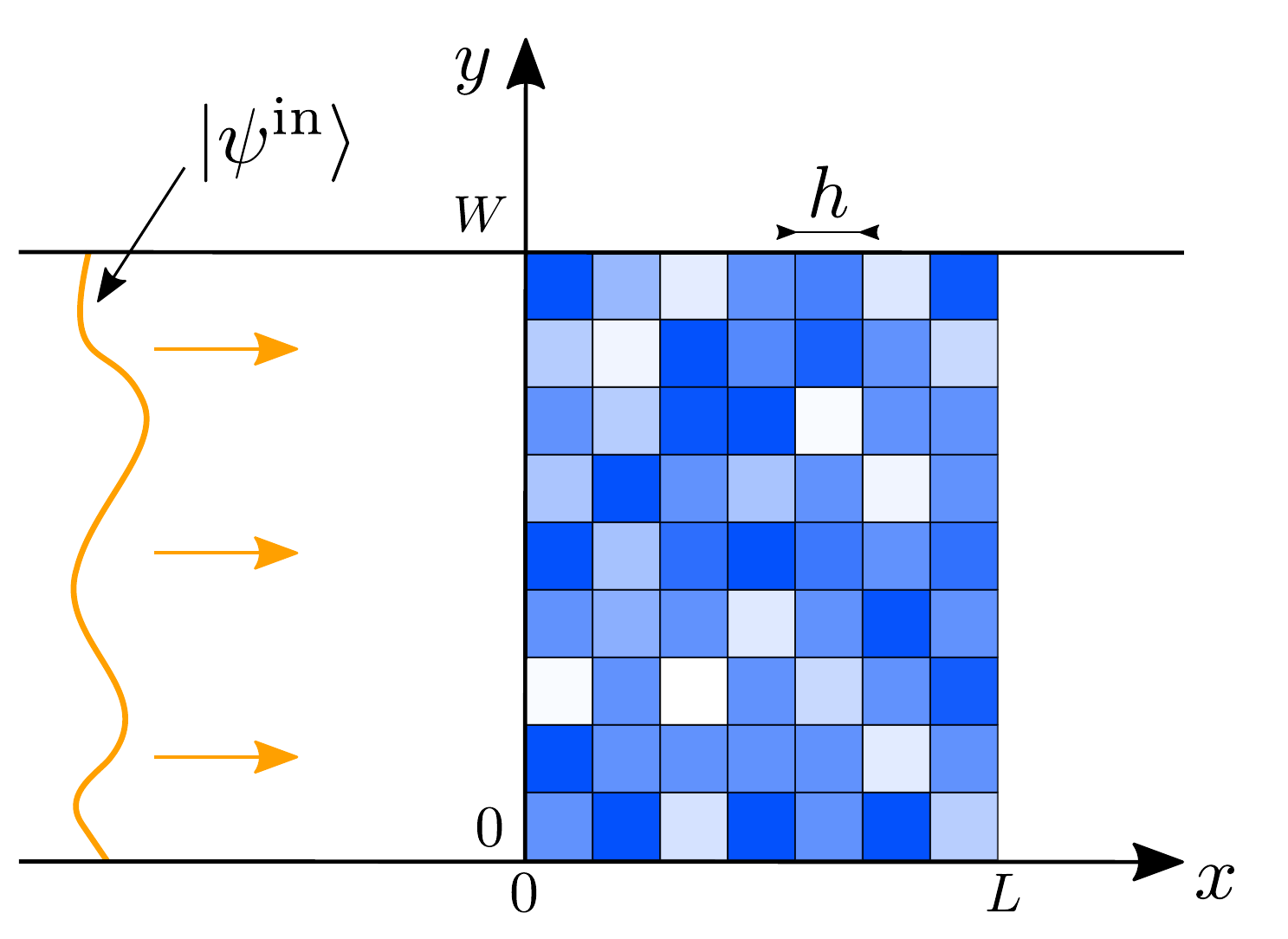}
\caption{ Model of the disordered system considered in this work. A disordered slab  of thickness $L$,  embedded in a multimode waveguide of transverse width $W$, is illuminated from the left by a wavefront $\ket{\psi^{\textrm{in}}}$ at frequency $\omega=ck$. The dielectric function is $\epsilon(\vec{r})=n_1^2+\delta\epsilon(\vec{r})$, with  $n_1=1.5$ and  $\delta\epsilon(\vec{r})=0$ in the empty waveguide, and  $n_1=1.5$ and $\delta\epsilon(\vec{r})$ uniformly distributed  in $[-a, a]$ for each discretization cell of the slab. The value of $a$ is adjusted to generate different scattering strengths $1/k\ell$ (see section III for the numerical evaluation of $\ell$). The cell size is $h=0.2/k$. 
}
\label{fig1_SI}
\end{figure}

Our starting point is Eq.~(2) of the main text (MT), which we express in terms of the normalized field
\be
\tilde{\psi}= \sqrt{\frac{\epsilon_0c}{2\phi^{\textrm{in}}}}\psi,
\label{NormField}
\ee
as
\begin{align}
U&=\frac{\phi^{\textrm{in}}}{2k}\int_{\mathcal{S}}
\textrm{d}\vec{r}\,\vec{n}. 
\!\left(\partial_\omega \tilde{\psi} \nabla \tilde{\psi}^*
-\tilde{\psi}^*\partial_\omega  \nabla \tilde{\psi}
\right)
\nonumber
\\
&=\frac{\phi^{\textrm{in}}}{2k} \left[m(L)-m(0) \right],
 \label{U}
\end{align}
where
\be
m(x)=\ps{\partial_x \tilde{\psi}(x)}{\partial_\omega\tilde{\psi}(x)}-\ps{\tilde{\psi}(x)}{\partial_\omega\partial_x\tilde{\psi}(x)},
\label{DefM}
\ee
and $\phi^{\textrm{in}}$ is the incoming flux through the front surface $S_1$,
\be
\phi^{\textrm{in}}=\frac{\epsilon_0c}{2k}\int_{S_1}\textrm{d}\vec{r}\,\vec{n}.\textrm{Im} \left(\psi^{\textrm{in}*}\nabla \psi^{\textrm{in}} \right).
\label{DefFlux}
\ee
We refer to Fig.\ref{fig1_SI} for a schematic view of the scattering problem.

In order to express $m(x)$ in terms of the reflection and transmission matrices of the slab, we  first  need to expand the field $\ket{\tilde{\psi}(x)}$ in the constant flux basis $\ket{\psi_\alpha}$, in which these matrices are defined. The elements $\ket{\psi_\alpha}$ are related to the transverse plane waves $\ket{\chi_\alpha}$ as
\be
\psi_\alpha(\vec{r})=\frac{\chi_\alpha(\vec{r})}{\sqrt{\mu_\alpha}}=\frac{e^{\textrm{i}\vec{q}_\alpha.\vec{r}}}{\sqrt{\mu_\alpha\mc{A}}},
\label{defBasisFlux}
\ee
with $\vec{q}_\alpha$ the transverse wave vector, $\mu_\alpha=\sqrt{1-q_\alpha^2/k^2}$ and $\mc{A}$ is the area covered by the surface $S_1$. In this basis, the input field reads $\psi^{\textrm{in}}(\vec{r},x)=\sum_\alpha c^{\textrm{in}}_\alpha\psi_\alpha(\vec{r})\varphi_\alpha(x)$, with $\varphi_\alpha(x)=e^{\textrm{i}k\mu_\alpha x}$, so that the incoming flux~\eqref{DefFlux} becomes
\be
\phi^{\textrm{in}}=\frac{\epsilon_0c}{2}\sum_\alpha\vert c^{\textrm{in}}_\alpha \vert^2.
\ee
The normalization~\eqref{NormField} implies that $\tilde{\psi}^{\textrm{in}}(\vec{r},x)=\sum_\alpha \tilde{c}^{\textrm{in}}_\alpha\psi_\alpha(\vec{r})\varphi_\alpha(x)$, with
\be
\sum_\alpha\vert \tilde{c}^{\textrm{in}}_\alpha \vert^2=1.
\label{NormPsi}
\ee
 According to the definition~\eqref{defBasisFlux}, the vector $\ket{\tilde{\psi}^{\textrm{in}}(x)}$  and the vector $\ket{\tilde{\psi}^{\textrm{in}}}$ of components $\tilde{c}^{\textrm{in}}_\alpha$ are related by
\be
\ket{\tilde{\psi}^{\textrm{in}}(x)}=P(x)\ket{\tilde{\psi}^{\textrm{in}}},
\ee 
where $P(x)$ is a diagonal matrix with elements
\be
 P_{\alpha\alpha}(x)=\frac{\varphi_\alpha(x)}{\sqrt{\mu_\alpha}}=\sqrt{\frac{k}{k_\alpha}}e^{\textrm{i}k_\alpha x}.
\label{DefP}
\ee
Thus, the field in front of the slab reads
\be
\ket{\tilde{\psi}(x)}=\left[P(x) + P(-x) \,r\right]\ket{\tilde{\psi}^{\textrm{in}}},
\label{FieldR}
\ee
and the transmitted field is
\be
\ket{\tilde{\psi}(x)}=P(x-L)\,t\ket{\tilde{\psi}^{\textrm{in}}}.
\label{FieldT}
\ee

According to the decompositions~\eqref{FieldR} and~\eqref{FieldT}, Eq.~\eqref{U}  can formally be written as the matrix element
\be
U=\phi^{\textrm{in}}\bra{\tilde{\psi}^{\textrm{in}}}Q_d\ket{\tilde{\psi}^{\textrm{in}}},
\label{UQd}
\ee
where
\be
Q_d=\frac{M(L)-M(0)}{2k}.
\label{DefQd}
\ee
The matrix $M(L)$ is found by inserting the decomposition~\eqref{FieldT} into Eq.~\eqref{DefM}. We get
\be
M(L)=-2\textrm{i}t^\dagger A \partial_\omega t+ t^\dagger Bt
\label{DefML}
\ee
with 
\begin{align}
A&=\textrm{Im}\left[P^{\dagger}(0)\partial_x P(0)\right],
\\
B&=\partial_x P^{\dagger}(0)\partial_\omega P(0)-P^{\dagger}(0)\partial_\omega\partial_x P(0).
\end{align}
With the help of Eq.~\eqref{DefP}, the elements of the diagonal matrices $A$ and $B$ can be expressed as
\begin{align}
A_{\alpha \alpha}&=k\frac{\textrm{Re}(k_\alpha)}{\vert k_\alpha \vert},
\\
B_{\alpha \alpha}&=k\left(\frac{\partial_\omega k_\alpha}{k_\alpha}\frac{\textrm{Im}k_\alpha}{\vert k_\alpha \vert} -\frac{\textrm{i Re}(k_\alpha)}{\omega\vert k_\alpha \vert} \right).
\end{align}
At this stage, it useful to split the matrices  $A$ and $B$ into two blocks corresponding to a separation between propagating channels ($k_\alpha=\sqrt{k^2-q_\alpha^2}>0$) and evanescent channels ($k_\alpha = \textrm{i} \kappa_\alpha$, with $\kappa_\alpha=\sqrt{q_\alpha^2-k^2}>0$), as $A=\left(\begin{smallmatrix}A^{pp} \ 0 \\ 0 \ A^{ee}\end{smallmatrix}\right)$ and $B=\left(\begin{smallmatrix}B^{pp} \ 0 \\ 0 \ B^{ee}\end{smallmatrix}\right)$, with
\begin{align}
A^{pp}_{\alpha \alpha}&=k \;\; \textrm{and} \;\; A^{ee}_{\alpha \alpha}=0,
\label{DefA}
\\
B^{pp}_{\alpha \alpha}&=-\frac{\textrm{i}k}{\omega} \;\; \textrm{and} \;\; B^{ee}_{\alpha \alpha}=k\frac{\partial_\omega \kappa_\alpha}{\kappa_\alpha}.
\label{DefB}
\end{align}

In the same way, the matrix $M(0)$ is found by inserting the decomposition~\eqref{FieldR} into Eq.~\eqref{DefM}. We  split $M(0)$ into three terms
\be
M(0)=B+M^{r}(0)+M^{c}(0).
\label{DefM0}
\ee
The second and third terms read
\begin{align}
M^{r}(0)&=2\textrm{i}r^\dagger A \partial_\omega r- r^\dagger Br,
\label{DefM02}
\\
M^{c}(0)&=2C_1 \partial_\omega r+C_2r-r^\dagger C_2,
\label{DefM03}
\end{align}
with
\begin{align}
C_1&=\textrm{Re}\left[P^{\dagger}(0)\partial_x P(0)\right],
\\
C_2&=\partial_x P^{\dagger}(0)\partial_\omega P(0)+P^{\dagger}(0)\partial_\omega\partial_x P(0).
\end{align}
The propagating and evanescent blocks of $C_1$ and $C_2$ have the following components
\begin{align}
C^{pp}_{1,\alpha \alpha}&=0 \;\; \textrm{and} \;\; C^{ee}_{1,\alpha \alpha}=-k.
\label{DefC1}
\\
C^{pp}_{2,\alpha \alpha}&=\textrm{i}k\frac{\partial_\omega k_\alpha}{k_\alpha} \;\; \textrm{and} \;\; C^{ee}_{2,\alpha \alpha}=-\frac{k}{\omega}.
\label{DefC2}
\end{align}

Hence, the general expression of the operator $Q_d$ is given by Eq.~\eqref{DefQd} with $M(L)$ defined in Eq.~\eqref{DefML} and $M(0)$ defined in Eqs.~\eqref{DefM0},~\eqref{DefM02}, and~\eqref{DefM03}. Their explicit expressions can be found by decomposing $t$ and $r$ as $t=\left(\begin{smallmatrix}t^{pp} \ t^{pe} \\ t^{ep} \ t^{ee}\end{smallmatrix}\right)$ and $r=\left(\begin{smallmatrix}r^{pp} \ r^{pe} \\ r^{ep} \ r^{ee}\end{smallmatrix}\right)$, and using  expressions of auxiliary matrices $A$, $B$, $C_1$, and $C_2$, given in Eqs.~\eqref{DefA},~\eqref{DefB},~\eqref{DefC1}, and~\eqref{DefC2}.  

 Let us now restrict the analysis to an input state $\ket{\tilde{\psi}^{\textrm{in}}}$ without evanescent component, for which we need to evaluate the $N\times N$ matrix $Q_d^{pp}$ only. It can be decomposed as
\be
Q_d^{pp}=Q+Q_a+Q_i+Q_e.
\label{QdDecompo}
\ee
The first term is the Wigner-Smith operator:
\begin{align}
Q&=-\textrm{i}\frac{t^{pp\dagger} A^{pp} \partial_\omega t^{pp}+r^{pp\dagger} A^{pp} \partial_\omega r^{pp}}{k}
\nonumber
\\
&=-\textrm{i}(t^{pp\dagger}  \partial_\omega t^{pp}+r^{pp\dagger}  \partial_\omega r^{pp}).
\end{align}
The second term of Eq.~\eqref{QdDecompo} is
\begin{align}
Q_a&=\frac{t^{pp\dagger} B^{pp}t^{pp}+r^{pp\dagger} B^{pp}r^{pp}-B^{pp}}{2k}
\nonumber
\\
& =\frac{\textrm{i}}{2\omega}\left(1-t^{pp\dagger} t^{pp} -r^{pp\dagger} r^{pp} \right),
\label{defQa}
\end{align}
which is zero in the absence of absorption. The third term of the decomposition~\eqref{QdDecompo} is due to $M^c(0)$ which corresponds to the interference  between incident and reflected fields:
\begin{align}
Q_i&=\frac{C_2^{pp}r^{pp}-r^{pp\dagger}C_2^{pp}}{2k}
\nonumber
\\
& =-\frac{\textrm{i}}{2}\left(Dr^{pp} -r^{pp\dagger} D \right),
\end{align}
where $D$ is a diagonal matrix with elements 
\be
D_{\alpha \alpha}=\frac{\partial_\omega k_\alpha}{k_\alpha}
=\frac{\omega}{\omega^2-c^2q_\alpha^2} \;\;\;(\alpha \le N).
\ee
The last term of Eq.~\eqref{QdDecompo} is due to scattering into evanescent channels of the empty waveguide: 
\begin{align}
Q_e&=\frac{t^{ep\dagger} B^{ee}t^{ep}+r^{ep\dagger} B^{ee}r^{ep}}{2k}
\nonumber
\\
& =\frac{1}{2} \left(t^{ep\dagger} D^{e}t^{ep}+r^{ep\dagger} D^{e}r^{ep}\right),
\end{align}
where 
$D^e$ is a diagonal matrix with elements 
\be
D^e_{\alpha \alpha}=\frac{\partial_\omega \kappa_\alpha}{\kappa_\alpha}
=\frac{\omega}{\omega^2-c^2q_\alpha^2} \;\;\;(\alpha > N).
\ee
Equations~\eqref{UQd} and~\eqref{QdDecompo}, with $\ket{\tilde{\psi}^{\textrm{in}}}$ satisfying the normalization condition~\eqref{NormPsi}, correspond to the formula (3) and (4) of the MT, for which we used slightly simplified notations. 

Finally, note that, in a waveguide of refractive index $n_1$ (as the one used in our simulations), the variables $k$ and $k_\alpha$ appearing in Eq.~\eqref{DefP} and following expressions must be multiplied by $n_1$.

\subsection{I.2 Case 2: arbitrary $\epsilon(\vec{r}, \omega)$} 

For arbitrary dielectric function $\epsilon$, simple manipulations of the wave equation lead to the following generalization of Eq.~(1) of the MT:
\begin{align}
&\frac{\epsilon(\vec{r})+\partial_\omega [\omega \epsilon(\vec{r})]}{2}
\vert\psi(\vec{r})\vert^2
+\textrm{i}\omega\textrm{Im}[\epsilon(\vec{r})]\psi(\vec{r})^*\partial_\omega\psi(\vec{r})
\nonumber
\\
&=\frac{c^2}{2\omega}
\mathbf{\nabla} .
\left(\partial_\omega \psi \nabla \psi^*
-\psi^*\partial_\omega  \nabla \psi
\right).
\end{align}
The right hand side of this relation is the same as in Eq.~(1) of the MT and has been computed in section I.1. The relation~\eqref{UQd}  is generalized as:
\be
W=\phi^{\textrm{in}}\bra{\tilde{\psi}^{\textrm{in}}}Q_d\ket{\tilde{\psi}^{\textrm{in}}},
\label{VQd}
\ee
with $Q_d$ given by Eq.~\eqref{QdDecompo}, and $W$ defined as
\begin{align}
W&=\frac{\epsilon_0 }{4}\int_{\mathcal{V}}\textrm{d}\vec{r}\, \{\epsilon(\vec{r})+\partial_\omega [\omega \epsilon(\vec{r})]\} \vert\psi(\vec{r})\vert^2 
\nonumber
\\
&+  \frac{\textrm{i}\omega\epsilon_0 }{2} \int_{\mathcal{V}}\textrm{d}\vec{r}\, \textrm{Im}[\epsilon(\vec{r})]\psi(\vec{r})^*\partial_\omega\psi(\vec{r}),
\label{DefW}
\end{align}
Note that $Q_a=\textrm{i}A/2\omega$, introduced in Eq.~\eqref{defQa}, is non-zero in the present situation, because the absorption operator $A$ is non-zero itself.

 The electromagnetic energy in the medium being given by the real part of the first term of Eq.~\eqref{DefW}~\cite{landau60, jackson73}, we considered the Hermitian part of $Q_d$ to optimize the dwell-time in the presence of absorbing resonators. The results are presented in Fig.~3 of the MT.

\subsection{I.3 Interpretation of formula (4) of the MT based on the continuity equation} 
It can be verified that the wave equation, with $\epsilon(\vec{r})$ real and independent of frequency, is consistent with the continuity equation
\be
\partial_tu(\vec{r}, t) +\nabla. \,\vec{j}(\vec{r}, t)=0,
\ee
where $u(\vec{r}, t)=\epsilon_0\epsilon(\vec{r})\,\textrm{Im}\left(\psi\partial_t\psi^*\right)/2ck$ and $\vec{j}(\vec{r}, t)=\epsilon_0c\,\textrm{Im}\left(\psi^*\nabla\psi\right)/2k$. The Fourier transforms of the energy density and the current, noted  $u(\vec{r}, \Omega)$ and  $j(\vec{r}, \Omega)$ respectively, obey
\begin{align}
\textrm{i}\Omega u(\vec{r}, \Omega)&=\nabla. \,\vec{j}(\vec{r}, \Omega),
\\
\textrm{i}\left[u(\vec{r}, \Omega)+ \Omega\partial_\Omega u(\vec{r}, \Omega) \right]&=\nabla. \,\partial_\Omega \vec{j}(\vec{r}, \Omega).
\end{align}
In the stationary regime, these relations reduce to 
\begin{align}
0&=\nabla. \,\vec{j}(\vec{r}, \Omega=0),
\label{Continuity1}
\\
\textrm{i}u(\vec{r}, \Omega =0)&=\nabla. \,\partial_\Omega \vec{j}(\vec{r}, \Omega =0).
\label{Continuity2}
\end{align}

By decomposing the total field as $\psi= \psi^{\textrm{in}} + \psi^s$ and the resulting current as $\vec{j}=\vec{j}^{\textrm{in}}+\vec{j}^{\textrm{s}}+\vec{j}^{c}$ [where $\vec{j}^{c}\sim\textrm{Im}\left(\psi^{\textrm{in}*}\nabla\psi^s\right)$ designates the cross-term], we obtain from Eq.~\eqref{Continuity1}, the optical theorem
\be
\oint_{\mathcal{S}}
\textrm{d}\vec{r}\,\vec{n}.\vec{j}^{\textrm{s}}(\vec{r}, \Omega=0)=-\oint_{\mathcal{S}}
\textrm{d}\vec{r}\,\vec{n}.\vec{j}^{\textrm{c}}(\vec{r}, \Omega=0).
\label{OpticalThm}
\ee

On the other hand, Eq.~\eqref{Continuity2} gives 
\be
\textrm{i} \int_\mathcal{V}\textrm{d}\vec{r}\, u(\vec{r}, \Omega =0)=\oint_{\mathcal{S}}\,\vec{n}.\partial_\Omega \vec{j}(\vec{r}, \Omega =0),
\label{Continuity3}
\ee
which turns out to be equivalent to Eq.~(2) of the MT. This allows us to interpret in a new way the different contributions of Eq.~(4) of the MT.  The field at the back surface is $\sim t \psi^{\textrm{in}}$, which gives rise to a single current contribution, $\partial_\Omega \vec{j}^t$, to Eq.~\eqref{Continuity3}. On the contrary, the field is   $\psi^{\textrm{in}} + r \psi^{\textrm{in}}$ at the front surface, which results in a current contribution $\partial_\Omega \vec{j}^{\textrm{in}} + \partial_\Omega \vec{j}^r + \partial_\Omega \vec{j}^c$ [where $\vec{j}^{c}\sim\textrm{Im}\left(\psi^{\textrm{in}*}\nabla\psi^r\right)$ designates the cross-term]. The terms $\partial_\Omega \vec{j}^t$ and $\partial_\Omega \vec{j}^r$ are responsible for the operator $Q+Q_e$ of Eq.~(4) of the MT, while the cross-term $\partial_\Omega \vec{j}^c$ leads to $Q_i$. This establishes a close connection between $Q_i$ and the interference term between the incident and scattered fields at the origin of the optical theorem~\eqref{OpticalThm}.

\begin{figure}[t]
\includegraphics[width=1\linewidth]{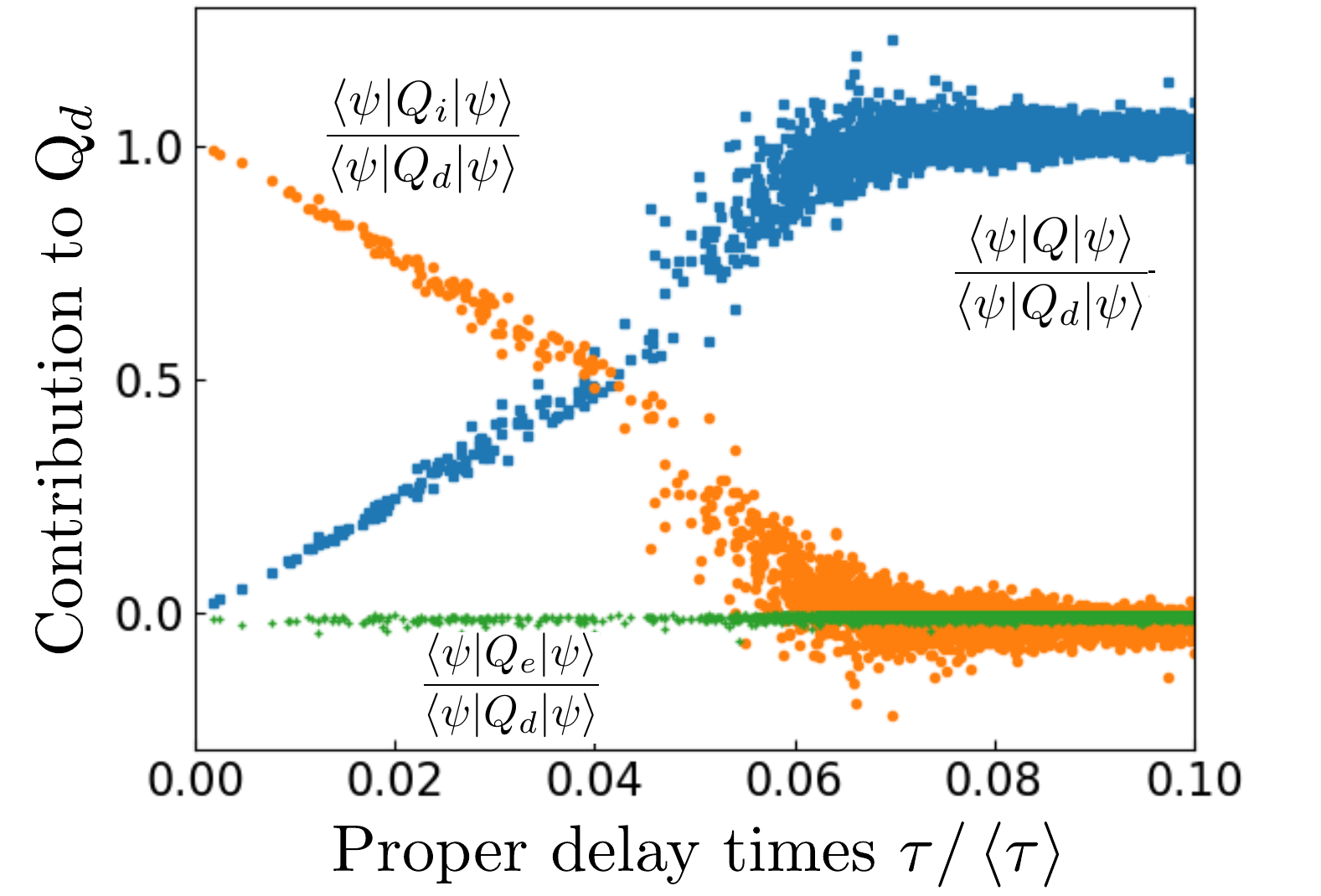}
\caption{Contributions of $Q$, $Q_i$ and $Q_e$ to $\bra{\psi}Q_d\ket{\psi}$ for eigenstates $\ket{\psi}$ of the Wigner-Smith matrix $Q$ associated to the eigenvalues $\tau$. No absorption is included, so that $Q_d=Q+Q_i+Q_e$ [see Eq.~\eqref{QdDecompo}]. Parameters of the simulations: $kL=300$, $kW=600$, $k\ell = 21.4$.
}
\label{fig2_SI}
\end{figure} 

\begin{figure*}[t]
\includegraphics[width=0.9\linewidth]{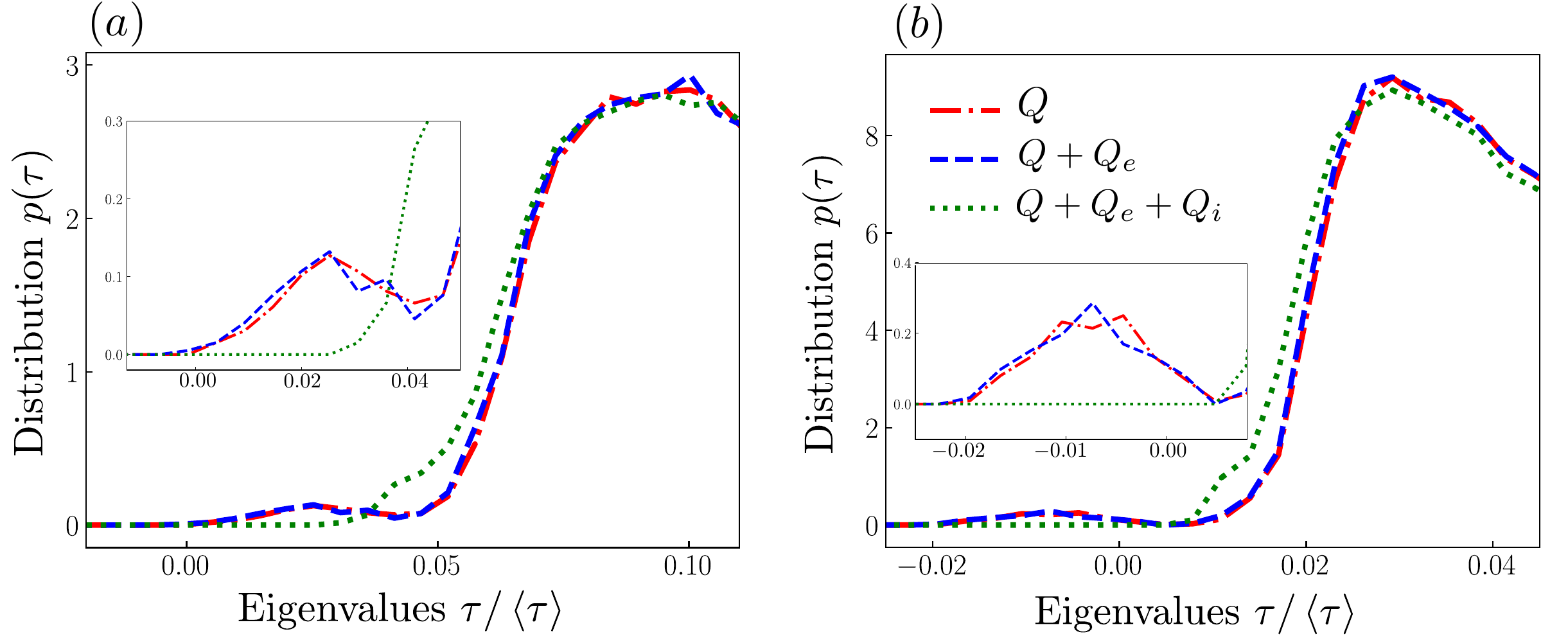}
\caption{Eigenvalue distribution $p(\tau)$ of the operators $Q$, $Q+Q_e$, and $Q_d=Q+Q_e+Q_i$ for two sets of parameters: (a) $kL=300$, $kW=600$, $k\ell = 21.4$, (b) $kL=450$,  $kW=600$, $k\ell = 9.5$. Only parts of the distributions that differ are shown. Insets are zooms of the distribution tails. 
}
\label{fig3_SI}
\end{figure*} 

\section{II. Contributions of $Q_i$ and $Q_e$ to $Q_d$}

\subsection{II. 1. Contribution of $Q_i$} 

As explained above, the term $Q_i$ in the decomposition~\eqref{QdDecompo} originates from the interference between the incident and reflected fields. As a result, the contribution $\bra{\psi}Q_i\ket{\psi}$ becomes appreciable for states $\ket{\psi}$ with large reflection. The proportion of such states is  important for optically thick materials, which reflect most of the light. In order to illustrate this effect, we  considered a disordered slab of optical thickness $L/\ell \simeq14 $, for which we evaluated the eigenstates $\ket{\psi_n}$ of $Q$ associated to the proper delay times $\tau_n= \bra{\psi_n}Q\ket{\psi_n}$, and computed the weights  $\bra{\psi_n}Q_i\ket{\psi_n}/\bra{\psi_n}Q_d\ket{\psi_n}$ and $\bra{\psi_n}Q\ket{\psi_n}/\bra{\psi_n}Q_d\ket{\psi_n}$. Results are presented in Fig.~\ref{fig2_SI}. It is seen that $\bra{\psi_n}Q_i\ket{\psi_n}$ is the dominant contribution to $\bra{\psi_n}Q_d\ket{\psi_n}$ at short proper delay time $\tau$. This is consistent with the fact that the reflection is reduced as $\tau$ is increased. In addition, the contribution of evanescent channels $\bra{\psi_n}Q_i\ket{\psi_n}$ is negligible for all  $\tau_n$.  

We also analyzed in details the differences between the eigenvalue distributions of the operators $Q$ and $Q_d$. Representative results are shown in Fig.~\ref{fig3_SI}. The distributions differ at small $\tau$ only. In particular, $Q$ presents a set of small eigenvalues that detach from the bulk of the distribution. These eigenvalues can in some case be negative [see Fig.~\ref{fig3_SI}(b)]. On the contrary,  the eigenvalues of $Q_d$ are all positive, as imposed by Eq.~\eqref{UQd}, and belong to the bulk of the distribution.  Finally, we note that the the contribution of $Q_e$ is negligible for the simulations presented in Fig.~\ref{fig3_SI}.

\subsection{II. 2. Contribution of $Q_e$} 

The results discussed in section II.1 suggest that the effect of $Q_e$ on the eigenvalue distribution $p(\tau)$ is negligible. This is the case except for specific choice of the frequency of the input wave. Indeed, if the later is  adjusted to obtain a particularly long decay length of the first evanescent channel, the role of $Q_e$ is strongly enhanced.  

\begin{figure}[t]
\includegraphics[width=0.8\linewidth]{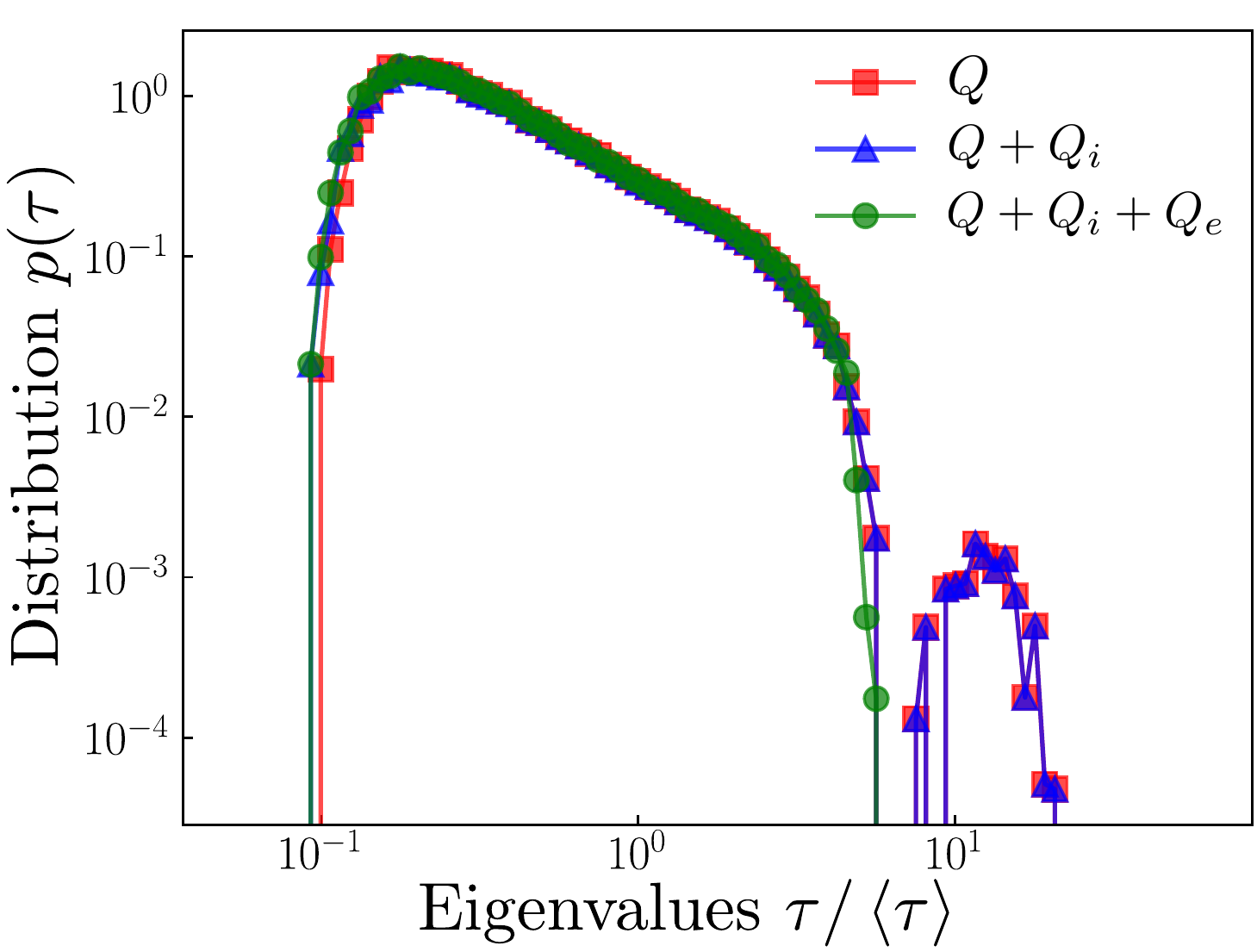}
\caption{Eigenvalue distribution $p(\tau)$ of the operators $Q$, $Q+Q_i$, and $Q_d=Q+Q_i+Q_e$, with $1-\delta=2.1\times 10^{-4}$ [see Eq.~\eqref{DefEvanescentLength} for the definition of $\delta$]. Parameters of the simulations: $kL=150$, $kW=231.6$, $k\ell = 21.4$, $kh=0.2$, $n_1=1.5$.
}
\label{fig4_SI}
\end{figure}

In a 2D waveguide of width $W$, the decay length of an evanescent channel $\alpha$ is $\ell_\alpha=1/\kappa_\alpha=[(\alpha\pi/W)^2-k^2]^{-1/2}$. 
By introducing the number of propagating channels $N=\lfloor kW/\pi \rfloor$, and the fractional part $\delta=kW/\pi -N>0$, the decay length of the first evanescent channel can be rewritten, in the limit $N\gg 1$, as
\be
\ell_{N+1}\simeq \frac{1}{k}\sqrt{\frac{N}{2(1-\delta)}}.
\label{DefEvanescentLength}
\ee
Hence, by choosing $\delta$ close to $1$ (this occurs when $kW/\pi$ approaches an integer value from below),  we can make $\ell_{N+1}$ as large as we wish. We computed the eigenvalue distribution $p(\tau)$ in this regime, and found differences between the spectra of $Q_d$ and $Q_s + Q_i$ at large $\tau$. As clearly seen in Fig.~\ref{fig4_SI}, a small fraction of large eigenvalues leaves the bulk of the distribution if $Q_e$ is not included, which leads to a large overestimate of the maximum  energy that we can store in the medium at the considered frequency. 

 Note that in a slab of index $n_1$, with discretization cells of size $h$, the value $kW/\pi$ appearing in the definition of $\delta$ must be replaced by $2W\textrm{arcsin}(n_1kh/2)/\pi h$.

\section{III. Analytic prediction for the dwell time eigenvalue distribution $p(\tau)$}
Our goal here is to find the distribution $p(\tau)=p(\gamma)/\gamma^2$ that minimizes the energy $\mc{H}$ defined in Eq.~(5) of the MT, under the constraint $\int \textrm{d} \gamma p(\gamma)/\gamma=\langle \tau \rangle$, where $\gamma=1/\tau$. In the large $N$ limit, we can perform a coarse-graining of $\mc{H}$ as 
\be
\mc{H}\simeq 2\tau_sN^2\int\textrm{d} \gamma  \gamma p(\gamma) -\frac{N^2}{2}\iint\textrm{d} \gamma \textrm{d} \gamma'p(\gamma)p(\gamma') \textrm{ln} \vert \gamma-\gamma'\vert,
\ee
where corrections of the order $N\textrm{ln} N$ have been neglected~\cite{dyson72, beenakker97}. The equilibrium distribution $p(\gamma)$ of the gas of eigenvalues is given by the extremum of this functional. The two constraints on the normalization and the first moment of $p(\tau)$  are taken into account by the introduction of two Lagrange multipliers $c_0$ and $c_1$. The condition $\partial_p\tilde{\mc{H}}=0$, with $\tilde{\mc{H}}=\mc{H}+c_0N^2\left[\int \textrm{d} \gamma p(\gamma)-1 \right]+ c_1N^2\left[\int \textrm{d} \gamma p(\gamma)/\gamma-\langle \tau \rangle \right]$, gives
\be
2\tau_s\gamma-\int\textrm{d} \gamma  p(\gamma') \textrm{ln} \vert \gamma-\gamma'\vert +\frac{c_1}{\gamma}+c_0=0.
\ee
By differentiating this equation with respect to $\gamma$, we get
\be
\textrm{P}\int\textrm{d}\gamma' \frac{p(\gamma')}{\gamma-\gamma'}=2\tau_s-\frac{c_1}{\gamma^2},
\ee
which we rewrite for the variable $\tilde{\gamma}=4\tau_s \gamma$ as 
\be
\textrm{P}\int\textrm{d}\tilde{\gamma}' \frac{p(\tilde{\gamma}')}{\tilde{\gamma}-\tilde{\gamma}'}=\frac{1}{2}-\frac{\mu}{\tilde{\gamma}^2},
\label{Fredholm}
\ee
where $\mu=4\tau_sc_1$. 
The solution $p(\tilde{\gamma})$, with finite support $[a, b]$, of the Fredholm  integral equation of the first kind~\eqref{Fredholm} is given by Tricomi's theorem~\cite{tricomi57}:
\begin{align}
&p(\tilde{\gamma})=\frac{1}{\pi^2}\frac{1}{\sqrt{(\tilde{\gamma}-a)(b-\tilde{\gamma})}}
\nonumber
\\
& \times \left[C+
\textrm{P}\int_a^b\textrm{d}\tilde{\gamma}' \frac{\sqrt{(\tilde{\gamma}'-a)(b-\tilde{\gamma}')}}{\tilde{\gamma}'-\tilde{\gamma}}
\left(\frac{1}{2}-\frac{\mu}{\tilde{\gamma}^2} \right)
\right],
\end{align} 
where $C$ is a constant to be determined. The values of $a$, $b$, $\mu$, and $C$ are fixed by the normalization of $p(\tilde{\gamma})$, and the conditions $p(\tilde{\gamma}=a)=0$, $p(\tilde{\gamma}=b)=0$, and $\int_a^b \textrm{d} \tilde{\gamma} p(\tilde{\gamma})/\tilde{\gamma}=\langle \tau \rangle/4\tau_s$. We note that this problem is formally equivalent to the one encountered in Ref.~\cite{grabsch16} in the study of the distribution of the Wigner time $\tau_W=\sum_{n=1}^N \tau_n/N$ in the limit $L \to \infty$. The solution $p(\tau)=4\tau_sp(\tilde{\gamma}=4\tau_s/\tau)/\tau^2$ takes the form given in Eq.~(6) of the MT:
\be
p(\tau)= \frac{2\tau_s}{\pi\tau^2}\sqrt{ \left( \alpha\frac{\tau}{\tau_s}-1\right) \left(1-\beta\frac{\tau}{\tau_s}\right)}\left(1+\gamma \frac{\tau}{\tau_s}\right),
\label{poftau}
\ee
where $\alpha=b/4$, $\beta=a/4$, and $\gamma$ are solutions of 
\begin{align}
&\frac{(1-\sqrt{\beta/\alpha})^2(3+2\sqrt{\beta/\alpha}+3\beta/\alpha)}{8\sqrt{\beta/\alpha}(1+\beta/\alpha)}=\frac{\av{\tau}}{4\tau_s},
\\
&\alpha=\frac{1+\beta/\alpha}{(1-\beta/\alpha)^2},
\\
&\gamma=\frac{2\beta/\alpha}{(1-\beta/\alpha)^2}.
\end{align}
The upper edge of the distribution is 
\be
\tau^{\textrm{max}}=\frac{\tau_s}{\beta}.
\label{TauMax1Edge}
\ee
Approximate expressions for $\alpha$, $\beta$, and $\gamma$ in the limit $\av{\tau}/\tau_s \gg 1$ are given in the MT.

 In addition, the two time scales $\av{\tau}$ and $\tau_s$ can be expressed in terms of the thickness $L$, the mean free path $\ell$, and the refractive index $n$. The mean time $\av{\tau}$  is given by
\be
\av{\tau}=\frac{\pi}{N}\av{\rho(\omega)},
\ee
where $\av{\rho(\omega)}$ is the mean density of states, related to the mean Green's function $G(\omega)$ of the wave equation by~\cite{akkermans07}
\be
\av{\rho(\omega)}=-\frac{2\omega}{\pi c^2}\av{\textrm{Im}\textrm{Tr}\,G(\omega)}.
\ee
In a 2D slab of thickness $L$ and width $W$, $\av{\textrm{Im}\textrm{Tr}\,G(\omega)}\simeq -LW/4$, and $N=\omega W/\pi c$. In an effective medium of refractive index $n$, the speed $c$ must be replaced by $v=c/n$ in previous expressions. We obtain
\be
\av{\tau}=\frac{\pi}{2}\frac{L}{v}.
\label{AvTau2D}
\ee
On the other hand, the scattering time $\tau_s$ appearing in Eq.~\eqref{poftau} is by construction proportional to $\ell/v$ [see Eq.~(5) of the MT]. What is unknown is the prefactor.  Our approach follows from a mapping on the problem considered in Ref.~\cite{bruce96}, which is itself an extension of the DMPK approach, developed to treat scattering in quasi-1D systems~\cite{beenakker97}. It is known that the scattering mean free path  used in this framework, $\ell^{\textrm{RMT}}$, differs from the scattering mean free path evaluated microscopically, $\ell$, by a numerical constant that depends on the dimensionality; in 2D, $\ell^{\textrm{RMT}}=\pi\ell/2$~\cite{beenakker97}. Since our mapping gives $\tau_s=\ell^{\textrm{RMT}}/v$, we used
\be
\tau_s=\frac{\pi}{2}\frac{\ell}{v}.
\label{Taus2D}
\ee

\begin{figure}[t]
\includegraphics[width=0.8\linewidth]{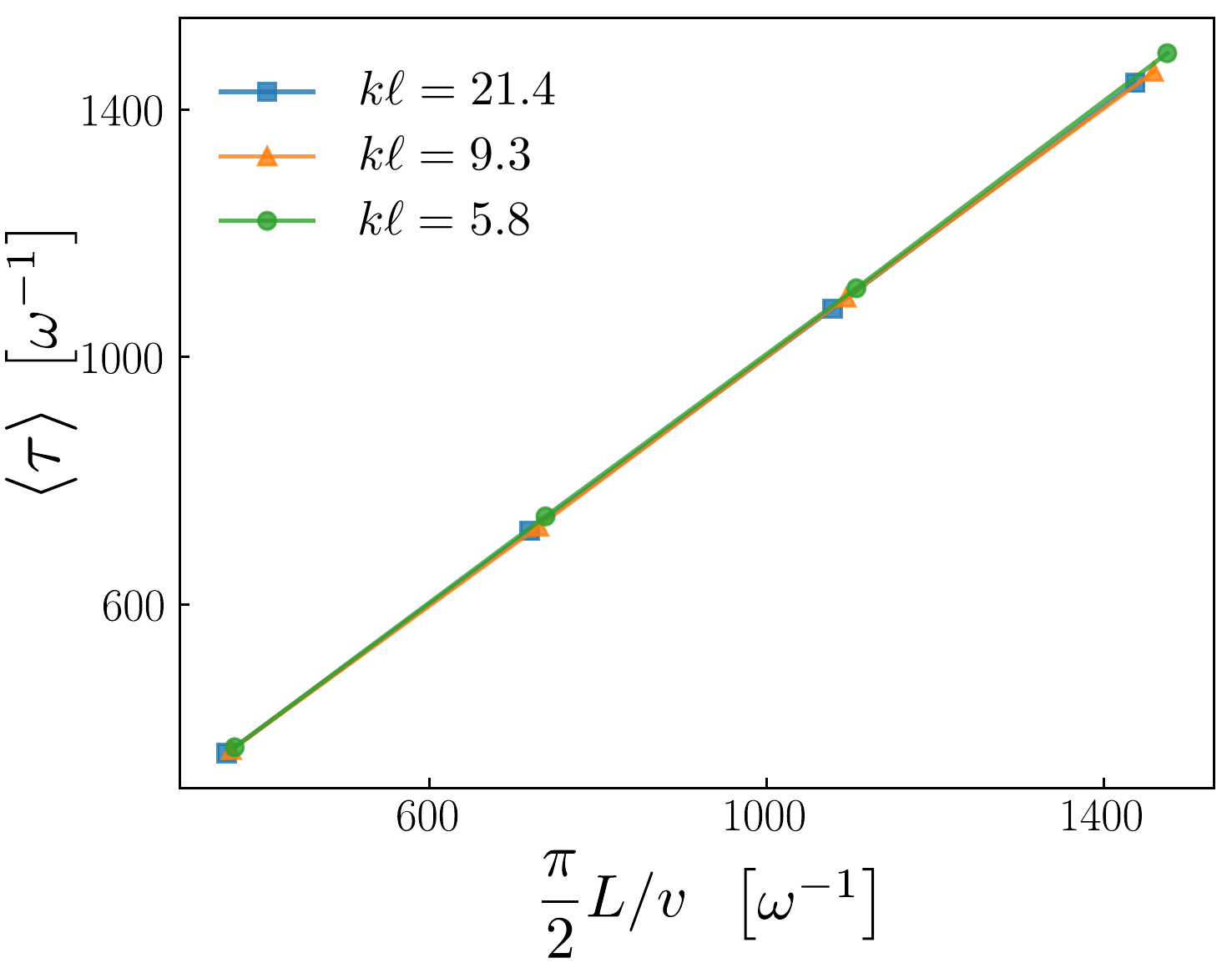}
\caption{Mean Wigner time $\av{\tau} = \av{\textrm{Tr}Q}/N$ as function of the theoretical result~\eqref{AvTau2D}, for three disorder strength $k\ell=21.4,\, 9.3,\, 5.8$, and four sample thickness $kL=150,\, 300,\, 450,\, 600$. 
}
\label{fig5_SI}
\end{figure}

In Fig.~1 and Fig.~2 of the MT, we compared the results of numerical simulations with the prediction~\eqref{poftau}, together with the expressions~\eqref{AvTau2D} and~\eqref{Taus2D} for $\av{\tau}$ and $\tau_s$. The value of the refractive index $n$ and the mean free path $\ell$ for a given disorder amplitude $a$ (see Fig.~\ref{fig1_SI} for the definition of $a$) are found by launching a plane wave perpendicular to the slab and fitting the intensity inside the disordered slab by $I(x)\sim \textrm{cos}(\omega x /v)^2e^{-x/\ell}$. The agreement between $\av{\tau} = \av{\textrm{Tr}Q}/N$ and Eq.~\eqref{AvTau2D}, using this procedure, is excellent, as shown in Fig.~\ref{fig5_SI}. 

\section{IV. Edge of the distribution for light injected from both sides of the sample}
In the scenario where light is injected from both sides of the samples, the result~\eqref{poftau} does not predict correctly the position of the right edge of the distribution $p(\tau)$. In this situation, the recent  model developed in Ref.~\cite{ossipov18} for electrons injected from both sides of the sample, turns out to be in good agreement with our numerical simulations, for a specific choice of the parameters of the model detailed below. Note that the mapping between electrons and light for the evaluation of the Wigner-Smith operator $Q$ is non-trivial, since the disordered potential for light depends on frequency, which is not the case for electrons.

 In the diffusive regime $L/\ell \gg1$ , the microscopic approach developed in Ref.~\cite{ossipov18} leads to the following prediction for the distribution $p(\tilde{\tau})$ of the variable $\tilde{\tau}=\tau/\tau_s$:
 \be
 p(\tilde{\tau})=-\frac{1}{\pi} \lim_{\epsilon \to 0^+} \, \textrm{Im} \,g(\tilde{\tau} + \textrm{i} \epsilon),
 \label{Solp}
 \ee
 where the resolvent $g(z)$ is given by
 \be
g(z)=\frac{1}{z^2}\left[z-2+2\sqrt{1-z+\frac{z^2}{z_0(z)^2}} \right].
 \label{Solg}
 \ee
The function $z_0(z)$ is solution of the implicit equation
\be
\frac{z_0}{2}\left[z_0\left[1+\textrm{cosh}\left(\frac{2r}{z_0}\right)\right]+2\textrm{sinh}\left(\frac{2r}{z_0}\right)\right]=z,
\label{Defz0}
\ee
where $r$ is a parameter proportional to the optical thickness (see the discussion below for its precise expression).

 Here, we are particularly interested in the right edge $\tau^{\textrm{max}}$ of the distribution  $p(\tau)$. In the large $N$ limit, the edges $z^*$ [$p(z^*)=0$] are such that the resolvent satisfies $\partial_zg(z^*)=\infty$.  This condition is fulfilled  for \be
\left.\frac{\textrm{d}z}{\textrm{d}z_0}\right\vert_{z^*}=0.
\ee
According to Eq.~\eqref{Defz0}, we thus find that the edges are given by
\be
z^*=\frac{z^*_0}{2}\left[z^*_0\left[1+\textrm{cosh}\left(\frac{2r}{z^*_0}\right)\right]+2\textrm{sinh}\left(\frac{2r}{z^*_0}\right)\right],
\label{SolzStar}
\ee
where $z_0^*$ is solution of the 
\be
(z_0^{*2}-2r)\textrm{cosh}\left(\frac{2r}{z^*_0}\right)+(1-r)z^*_0\textrm{sinh}\left(\frac{2r}{z^*_0}\right)+z_0^{*2}=0.
\label{Defz0Star}
\ee
In the limit $r\gg1$, we can search  an approximate solution for $z_0^*$ in the form  
$z_0^*\simeq \xi(r+1)$, which applies for the right edge of the distribution. By inserting this trial function into Eq.~\eqref{Defz0Star}, we find that $\xi$ is solution of the transcendental equation
\be
\xi+\xi\textrm{cosh}(2/\xi)-\textrm{sinh}(2/\xi)=0,
\ee
which gives $\xi\simeq0.83$. With this expansion for $z^*_0$, we now approximate Eq.~\eqref{SolzStar} as
\be
z^*\simeq 4 \zeta \left[ (r+1)^2-2\right]-1,
\ee 
where $\zeta=(\xi/8)\textrm{sinh}(2/\xi)\simeq 0.57$. Hence, at large optical thickness, the upper edge of the distribution $p(\tau)$ is well approximated by 
\be
\tau^{\textrm{max}}\simeq \tau_s  \left[4\zeta \left( r^2+2r-1\right)-1\right].
\ee

In Ref.~\cite{ossipov18}, the two parameters $\tau_s$ and $r$ are defined, in terms of the mean free path $\ell$, as $\tau_s=\ell/v$ and $r=\pi L/2\ell$ for 2D scattering systems. Since the  theoretical prediction for $p(\tilde{\tau})$
 implies $\av{\tilde{\tau}}=2r$, this would mean that $\av{\tilde{\tau}}=\pi L/v$, which is incompatible with Eq.~\eqref{AvTau2D}. In order to get a good agreement between our simulations and the prediction for $p(\tilde{\tau})$, we used instead $\tau_s=\pi\ell/2v$  and $r=L/2\ell$.

In Fig.~2(b) of the MT, the theoretical predictions~\eqref{TauMax1Edge} and~\eqref{SolzStar} for the upper edge of the distribution are compared with the results of numerical simulations. In order to evaluate $\tau^{\textrm{max}}$ in our simulations, we built  the  cumulative distribution functions $C(x)=\int _0^x\textrm{d}\tau p(\tau)$ from the numerical distributions $p(\tau)$, and solved $C(\tau^{\textrm{max}})=1-\epsilon$ with $\epsilon=5\times10^{-4}$.  

%